\documentstyle[prl,aps,epsfig,rotate]{revtex}

\begin{document}

\newcommand{\be}{\begin{equation}}
\newcommand{\ee}{\end{equation}}
\newcommand{\bea}{\begin{eqnarray}}
\newcommand{\eea}{\end{eqnarray}}
\newcommand{\vp}{\varphi}
\newcommand{\kmps}{\;{\rm km/s}}

\draft

\title{Directional Sensitivity, WIMP Detection, and the Galactic Halo}


\author{Craig J. Copi$^1$, Junseong Heo$^2$ and Lawrence M. Krauss$^{1,3}$
}
\address{$^1$Department of Physics, $^3$Department of Astronomy \\
Case Western Reserve University, 
10900 Euclid Ave., Cleveland OH 44106-7079}
\address{$^2$Department of
Physics, Yale University, New Haven, CT~06520-8120}

\date{\today}

\wideabs{

\maketitle

  
\begin{abstract}
\widetext

Distinguishing the signals due to scattering of WIMP
dark matter off of nuclear targets from those due to background noise is a
major challenge. The Earth's motion relative to the galactic halo should
produce halo-dependent seasonal modulation in the event rate, but it
also should produce an angular signal that is both far stronger and less
ambiguous. Distinct patterns in the recoil
spectrum can reflect the details of the galactic halo. We derive a new
formalism to calculate angular event rates, and present the predicted angular
signal for a variety of halo models and calculate the number of events needed
to distinguish a dark matter signal from an isotropic background.

\end{abstract}


}

\narrowtext

The direct experimental detection of WIMP dark matter would have profound
implications for both particle physics and cosmology.  Unfortunately, the
signature, excess energy deposits due to scattering on nuclei, is not easily
distinguishable from various radioactive background signals.  A significant
number of events might be needed before a definitive observation is
believable.  Moreover, the recoil spectrum alone is largely independent of
many features of the galactic halo distribution, so that our ability to
distinguish between galactic halo models will be limited by such observations.

In order to distinguish and isolate the WIMP signal, new signatures are of
great interest.  It was recognized quite early on that the Earth's motion
through the galaxy should induce both a seasonal variation in the overall
event rate and an overall forward-backward asymmetry in any directional
signal\cite{sadoulet,spergel}.  The seasonal modulation is problematic,
however.  Not only is it very small, of the order of $(2v_\oplus/v_{\rm
  halo})^2 \approx 0.03-0.05$, but as has become clear as a result of recent
claimed dark matter detections, the backgrounds themselves are likely to have
seasonal modulation, via modulations in the such things as cosmogenic
production of radon, etc.

The directional signature is, a priori, preferable in almost every way.
The forward backward asymmetry can be large, 
$O(v_\odot/v_{\rm halo}) \approx 1$, and backgrounds are unlikely to
reproduce this signature.
Nevertheless, in spite of its intrinsic interest and probably because
detectors with angular sensitivity are not yet on line, no systematic study of
possible angular signals has yet been performed.  But because of the
difficulty of building such detectors, it is worthwhile to examine in advance
what might be possible to learn with directional sensitivity.

Indeed, one of the virtues of this approach is also one of its drawbacks.
Angular resolution would provide sensitivity to the detailed features of the
galactic halo WIMP distribution.  However, because we do not know the nature
of this distribution in advance, we cannot assume its form in advance in order
to search for a signal.  In fact, one might imagine that
existing uncertainties are so great that one might not be able to obtain any
unambiguous limits on the basis of any lack of asymmetry in the observed
signals.  It is clear that the estimates which have previously
been carried out, for a spherically symmetric isothermal halo, are not
sufficient to determine in general how many events may be needed to distinguish
signals from backgrounds.  In the past decade we have learned that the galaxy
has large asphericities.  It has also has been argued that the halo
dark matter distribution may not resemble an isothermal gas at all
\cite{sikivie}.

In order to address these features, we have developed a new formalism which
allows one to calculate the full differential event rate in detectors, as a
function of both energy and angle, for any incident WIMP distribution, without
assumptions of spherical symmetry or even cylindrical symmetry.  In addition,
the full motion of the earth around the Sun is taken into account, not merely
the component of the Earth's velocity tangential to the Sun's galactic
velocity. While the perpendicular component is largely irrelvent for
estimating angular dependent effects if the incident halo is spherically
symmetric, this is not the case if this assumption is relaxed.  We report here
on our general results, leaving a detailed derivation to a later publication.
In addition, we summarize the main results from our first application of this
formalism.  We have generated Monte Carlo distributions for the full
spectrum of reported halo models, in order to perform a statistical analysis
to determine to what extent any directional signal might unambiguously be
distinguishable from an isotropric background.  Our results are strking.  We
find that, independent of the halo model, for energy thresholds and quenching
appropriate to the present generation of WIMP detectors, if fine scale angular
resolution is possible, fewer than 25 events will be needed to distinguish the
halo signal from a uniform background if the signal to noise ratio is large,
and fewer than 50 events would be required if the signal to noise ratio is
unity. On the other hand, if only a forward-backward asymmetry is discernable,
then between 500-3000 events may be required to distinguish a signal,
depending upon the actual halo distribution. In future work we will explore how
one might distinguish between different halo models, and how varying angular
and energy resolution and WIMP cross sections will effect these estimates.

Consider a WIMP of mass $m_\chi$ with its direction specified by the two
angles $(\alpha, \beta)$ incident on a fixed nucleus.  The WIMP elastically
scatters off the fixed nucleus and the scattering event can be described by
the two angles $(\theta^{*}, \xi)$ defined in the {\it center-of-mass\/}
frame.  The final angle set $(\gamma, \phi)$ describes the direction of the
{\it recoiled\/} nucleus in the lab.  Note that only the direction of the
recoiled nucleus is visible to the detector.  Neither the incident nor the
scattered direction of WIMP can be observed.  The probability of an incident
direction $(\alpha, \beta)$ is given by the halo model which defines the WIMP
distribution function.  Once we relate the three set of angles discussed above
using kinematics, we have only to find the proper Jacobian transformations in
order to present the final event rate as a function of the visible angles
$(\gamma, \phi)$.  This is simplified by the introduction of a single function
$J(\alpha,\beta;\gamma,\phi)$ defined as a scalar product of two unit vectors
on the sphere,
\begin{equation}
J(\alpha,\beta;\gamma,\phi)= [{\rm cos}\gamma\,{\rm cos}\alpha\> + 
\> {\rm sin}\gamma\,{\rm sin}\alpha\,{\rm cos}(\phi - \beta)].
\label{eqn:jacobian}
\end{equation}
This observation leads to the general angle dependent event rates $dR/d\Omega$
and $d^2\!R/{dQ\,d\Omega}$ for arbitrary distribution functions.  Here $Q (v,
J)$ is the energy transferred to the nucleus during the collision with the
WIMP and the $\Omega$ is the angle along which the target nucleus recoils.
For this work we focus on the angular distribution
\begin{eqnarray}
&&{{dR}\over {d\Omega_{\gamma,\phi}}}={{\sigma_{o} \rho_{\chi}} \over {\pi
m_{n} m_{\chi}}} \times \nonumber \\ &&\int_{v_{\rm min}}^{v_{\rm esc}} v^3 dv
F^{2}(Q) {\int d\Omega_{\alpha,\beta}}
f(v,\alpha,\beta)\,J(\alpha,\beta;\gamma,\phi)\Theta(J)
\end{eqnarray}
where $v_{\rm min}$ is related to the threshold energy, $\sigma_0$ is
the low energy WIMP-nucleus scattering
cross section, $v_{\rm esc} =
650\;{\rm km/s}$ is the
escape velocity of the galaxy, $\rho_\chi$ is the local dark matter halo
density, $\Theta(J)$ is a unit step function, $m_n$ is the mass of the target
nucleus, and $F(Q)$ a the form factor suppression for scalar interactions when
the WIMP mass couples to the quantum numbers of the entire target nucleus. In
this work, we take a simple exponential form for this form factor $F^2(Q)
\propto \exp(-Q/Q_0)$, where $ Q_0 = {3 \over {2m_n R_0^2}} $
and $m_n$ is the mass of the target nucleus and $R_0 = 0.3 + 0.91
\sqrt[3]{m_n}$ is the radius of the nucleus (in femtometers when $m_n$ is in GeV).

This formulation of the differential cross section is general enough to
accomodate any incident WIMP distribution function $f(v,\alpha,\beta)$.
However, when we evaluate these event rates, extra care is needed in the lower
limit of the velocity integration defined by the threshold energy of the
detector.  Unlike the angle independent analysis where the lower limit is
fixed once we have the specific threshold or the expected transfered energy,
the limit here varies as the incident direction changes for a fixed detector
position.  This is the main source of the variation of the event rate as a
function of angle in the angle dependent spectrum.  This change is also taken
care of with a division of the lower limit by the same
Jacobian~(\ref{eqn:jacobian}). The angular distributions shown have been 
averaged in 5 day bins over the Earth's motion for one year.

We next turn to the galactic halo models. Our desire here is to span the range
of realistic possibilities that have been explored in the literature, in order
to get a good idea of the existing uncertainties.  This might be an important
consideration for experimentalists who may devote considerable time to
attempting to features in the WIMP distribution that might not be guaranteed
to exist.  We consider the following: isothermal models, ``Evans"
models~\cite{evans} which are axisymmetric and allow for flattening.
(previously been  studied in the context of annular
modulations~\cite{kamionkowski}), co-rotating models with a net
angular momentum for the halo  \cite{lynden-bell}, and finally the most exteme
departure from an isothermal model, using phase space flows dominated by
infall into the galaxy for non-dissipative WIMPS, the ``Caustic"
model \cite{sikivie}.  

For the isothermal model we will consider three velocity dispersions,
$\left<v^2\right> = 3v_0^2/2$ where $v_0 = 150 \kmps$, $220\kmps$, and
$300\kmps$.  For the Evans model the important parameter is the flattening,
$q$.  We consider the values $q= 1$ (cored isothermal distribution), $0.85$,
and $1/\sqrt{2}$ (maximal flattening).  For the Caustic model the parameters
for the dark matter streams are given in table [1]
\cite{sikiviepriv}.  If we
assume the local density of dark matter is $0.52\;{\rm GeV}\;{\rm cm^{-3}}$,
appropriate to fits to the galactic rotation curve in this
model then  by summing the density in each caustic peak we find  that 60\%
of the local density is due to caustics.  The other 40\% we assume comes from
an isothermal distribution with
$v_0=220\kmps$.

To model a rotating halo we follow a 
standard prescription~\cite{lynden-bell}.
Let $f_+ (v,\alpha,\beta) = f(v,\alpha,\beta)$ when $0\le\alpha\le\pi/2$
and 0 otherwise.  Similarly let
$f_- (v,\alpha,\beta) = f(v,\alpha,\beta)$ when
$\pi/2\le\alpha\le\pi$ and 0 otherwise.  The distribution for a 
rotating halo is then given by
$  \tilde f = (1+\kappa) f_+ + (1-\kappa) f_- $.
Note that $\left< v^2 \tilde f \right> = \left< v^2 f \right>$.  The
average velocity of the halo matches the velocity of the Sun when
$\kappa = \sqrt{\pi}/2$.

With these analytic distribution functions, we can utilize our general
formalism to calculate the angular event rates, forward backward asymmetries,
and annual modulations expected in detectors.  A central issue mentioned
above, and the one we focus on here, is whether existing
uncertainties in galactic halo models might invalidate the approach of
utilizing angular resolution to extract a WIMP signature from an otherwise
isotropic background.

To display event rate distributions we choose a set of detector
parameters.We use germanium ($m_{\rm Ge} = 73\;{\rm GeV}$) as
the target nucleus. Note that germanium has a quenching factor of $0.25$,
thus only 25\% of the incident energy gets transfered to the recoiling
nucleus.  This is  important for determining detector thresholds.  In the
figures an assumed WIMP mass of $60\;{\rm GeV}$ was used, and thresholds 
incorporate this quenching factor.  We did not assume a specific particle
cross section, detector size nor efficiency. Instead we calculate the number of
events required for identification of a signal.  Based on this number and the
local halo density the detector size and efficiency can be determined for a
given cross section.

\begin{figure}[tbp]
  \leavevmode\center{\epsfig{figure=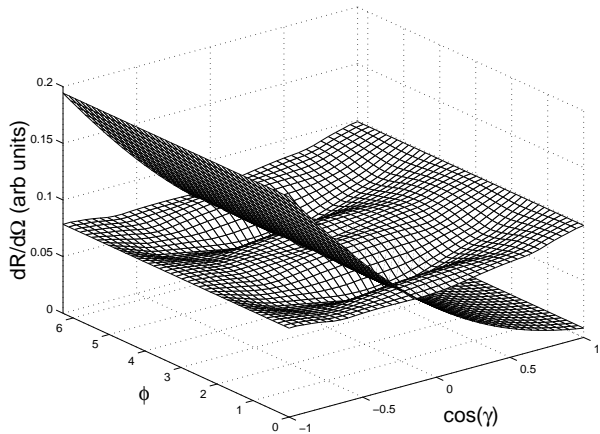,width=8cm}}
\caption{Angular event rate distribution for an isothermal model with
$v_0=220\kmps$, and for a caustic infall model with parameters described in
the text.  The isothermal
  model falls off exponentially in the forward direction while the caustic
model peaks slightly in the forward direction but 
is more isotropic due to the presence of both caustic and
isothermal components in the model.}
\label{fig:dRdO-isothermal}
\end{figure}

We employ a maximum likelihood analysis, along with Monte Carlo generation of
sample scattering distributions, as follows. We define a likelihood function
\begin{equation}
{\cal L} \equiv \prod_{i=1}^{N_e} P\left(\gamma_i,\phi_i\right),
\end{equation}
where $N_e$ is the total number of events and $P\left (\gamma_i, \phi_i
\right)$ is the probability of a nuclear recoil in the $\gamma_i$, $\phi_i$
direction based on a particular model (e.g.\ an isothermal distribution).  At
the 95\% confidence limit when
$\log {\cal L}_{dR/d\Omega} - \log {\cal L_{\rm flat}} < 1 $  the two
distributions are indistinguishable.  We generate 10,000 data sets for each
$N_e$ and demand that the log-likelihood condition
is satisfied less than 5\% of the time. The smallest $N_e$ for which this
occurs is the minimum number of events required to get a 95\% detection 95\%
of the time.

With the above procedure we can determine the number of events required to
distinguish the signal from a flat distribution, both for a pure signal and when
the signal-to-noise ratio,
$\rm S/N$, is one.  For
${\rm S/N} = 1$ we replace $P$ with
$  P_{\rm total} = \lambda P_{\rm flat} + (1-\lambda) P_{dR/d\Omega} $
where
$ \lambda = \frac1{1+\rm S/N}$.
We can perform a similar analysis for  the forward-to-backward
asymmetry.  In this case the probability function $P$ is given by a binomial
distribution.

\begin{figure}[tbp]
  \leavevmode\center{\rotate[r]{\epsfig{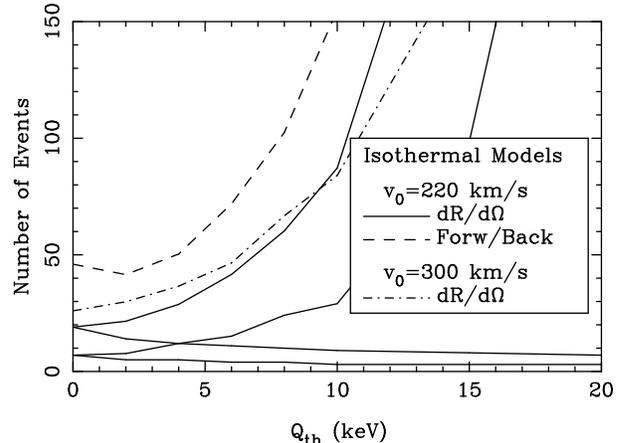}}}
\caption{
Required number of signal events for isothermal halos against the null
hypothesis.  Solid lines correspond to the $v_0=220\kmps$ and dashed lines to
$v_0=300\kmps$ . For the former, both $ ${\rm S/N}$=1$ case (upper) and
no noise cases are shown as are both the number of signal
events as a function of threshold and as the equivalent number
of events required for zero threshold.  For
the other case only the latter curve, for $ ${\rm S/N}$=1$ is shown. This
latter curve is also displayed when only the forward-backward asymmetry
is used to probe a
$v_0=220\kmps$ isothermal halo. }
\label{fig:isothermal}
\end{figure}

The results as a function of the threshold for an isothermal halo with 
velocity dispersions of $220\;$km/s and $300\;$km/s are given in
Figure~\ref{fig:isothermal} and for a
forward-backward asymmetry measurement for the $220\;$km/s halo. In this
figure, for the $220\;$km/s halo case, the bottom sets of lines give the number
of signal events required as a function of the threshold.  Fewer events are
required to distinguish a halo from a flat distribution in this case at higher
threshold because the anisotropy of the distribution becomes more dramatic as
higher scattering energies, and thus higher incident velocities relative to the
Earth, are required.  However, while fewer events are required as the
threshold is increased, even fewer events are expected, for a given detector
size, as the threshold is increased.  Thus, one does not win by simply raising
the theshold.  To make this clear, the upper curves show the number of events
(proportional to the size of the detector) that would be needed to be observed
in a zero threshold detector in order to obtain the requisite number of events
at each non-zero threshold.  This curve is a monotonically increasing function
of threshold.  Clearly, the lower the threshold, then, the smaller the
detector, and the smaller the total number of scatterings required in the
detector in order to distinguish the distribution from a flat distribution.

The upper two sets of curves for the $220\;$km/s halo case in the figure
represent the same requirement, but now in 
the case when the signal to noise ratio, ${\rm S/N}=1$ at each value
of the threshold.  The number of events plotted are the signal events.
The noise events would lead to twice as many total events 
actually being detected.  For the $300\;$km/s case, 
the forward-backward asymmetry, and the the co-rotating halo, Caustic, and 
Evans models (fig. 3)  only the latter ${\rm S/N}=1$, zero
threshold-equivalent number of events are shown.

\begin{figure}[tbp]
  \leavevmode\center{\rotate[r]{\epsfig{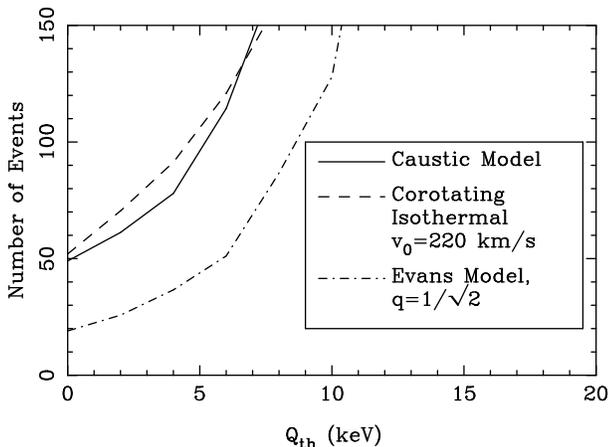}}}
\caption{
Required number of signal events for the Caustic model, a co-rotating, and an
Evans model (with $q=1/\sqrt2$) against the null
hypothesis, as in figure 2.  Only ${\rm S/N}=1$, zero
threshold-equivalent number of events are shown for these models.
}
\label{fig:caustic}
\end{figure}

Perhaps the most suprising result of our analysis is that, independent of the
existing uncertainties in the halo distribution, that less than 50 detected
events  for a signal to noise ratio of unity would be required in order to
unambiguously distinguish a halo induced signal from that due to a flat
background, assuming fine grained angular resolution.  If, however, only
forward/backward sensitivity were available, the requisite number of events
increases dramatically for non-isothermal distributions. This number rises to
500 for the co-rotating case, and 3000 for the Caustic case.  Next, we note
the suprising result that the Caustic distribution, which in principle is the
most distinctive halo distribution of all, requires the greatest number of
events in order to be distinguished from a flat distribution.  There are two
reason for this.  First, because this distribution is dominated by
infall components, the induced anisotropy in the pure cautic distribution is
in the {\it opposite} direction to that predicted in a standard isothermal
model. Adding a isothermal component then tends to cancel the anisotropy
induced by the Caustic flows.  In addition, because there is no exponential
tail for the caustic, the aniostropy tends to rise linearly in the forward
instead of exponentially (in the backward direction) as in the pure isothermal
models. 

For the same reason it is clear that the Caustic distribution
should be more easily distinguishable from these other distributions than it is
from a flat background.  The formalism we have presented here will allow
quantitative estimates for any observables to be derived for any incident halo
distribution.  Indeed, in future work we plan to
investigate how many events will be required in detectors in order to
distinguish between the various halo models, as well as incorporating 
additional detector features such angular resolution and energy detection
uncertainties as well as the theoretical distinctions between spin-dependent
and spin-independent scattering cross sections.  Our initial results, however
are encouraging.  Independent of halo uncertainties, good angular resolution
would allow an anisotropic halo signal to be differentiated from a flat
background signal with a fraction of the number of events that would be needed
to probe for annual modulation, or other expected spectral features due to a
WIMP halo in the galaxy. If a forward-backward asymmetry is all that can be
detected, the number of events required is model dependent, and can be quite
large, however.

 This work was supported by the DOE.

\begin{table}
\caption{Velocity flows of dark matter from the Caustic model (for
$h=0.75$, $\epsilon =0.28$, $j_{\rm max}=0.25$ Sikivie model).
Velocities are given in the rest frame of the galaxy. }
\label{tab:caustic}
\begin{tabular}{ccccc}
Flow 	&
  $\rho$\tablenotemark[1]
   & $v_x$\tablenotemark[2]
  & $v_y$\tablenotemark[2] & $v_z$\tablenotemark[2] \\
\tableline 1 & $0.4$ & $140$ & $0$ & $\pm600$ \\
2 & $0.9$ & $250$ & $0$ & $\pm500$ \\
3 & $2.0$ & $350$ & $0$ & $\pm395$ \\
4 & $6.1$ & $440$ & $0$ & $\pm240$ \\
5 & $9.6$ & $440$ & $\pm190$ & $0$ \\
6 & $3.0$ & $355$ & $\pm290$ & $0$ \\
7 & $1.9$ & $295$ & $\pm330$ & $0$ \\
8 & $1.4$ & $250$ & $\pm350$ & $0$ \\
9 & $1.0$ & $215$ & $\pm355$ & $0$ \\
10 & $1.1$ & $190$ & $\pm355$ & $0$ \\
\end{tabular}
\tablenotetext[1]{In units of  $10^{-26}\;{\rm g}\;{\rm
cm^{-3}}$}
\tablenotetext[2]{velocities shown are in $\rm
km\;s^{-1}$}
\end{table}

\end{document}